\documentclass[a4paper, 10pt, conference]{ieeeconf}      

\IEEEoverridecommandlockouts                              

\usepackage{graphics} 
\usepackage{epsfig} 
\usepackage{mathptmx} 
\usepackage{times} 
\usepackage{amsmath} 
\usepackage{amssymb}  
\usepackage{algorithm,algorithmic}
\usepackage{multicol}
\usepackage{multirow}
\usepackage{bm}
\usepackage{cite}
\usepackage{subcaption}

\usepackage{amsthm}
\usepackage{comment}

\newtheorem{rem}{Remark}

\title{\LARGE \bf
Adaptive FRIT-based Recursive Robust Controller Design Using Forgetting Factors
}

\author{Satoshi Tsuruhara$^{1}$ and Kazuhisa Ito$^{2}$
\thanks{*This work was supported by JSPS Grant-in-Aid for JSPS Fellows Grant Number JP23KJ1923.}
\thanks{$^{1}$Satoshi Tsuruhara with the Graduate School of Engineering and Science, Functional Control Systems, Shibaura Institute of Technology, $307$ Fukasaku, Minuma, Saitama $3378570$, Japan, and also is a research fellow with Japan Society for the Promotion of Science, Tokyo $1020083$, Japan
        {\tt\small nb23110@shibaura-it.ac.jp}}%
\thanks{$^{2}$Kazuhisa Ito with Department of Machinery and Control Systems,
         Shibaura Institute of Technology,
         307 Fukasaku, Minuma, Saitama 3378570, Japan.
        {\tt\small kazu-ito@shibaura-it.ac.jp}}%
}

\begin{document}

\maketitle
\thispagestyle{empty}
\pagestyle{empty}

\begin{abstract}
Adaptive FRIT (A-FRIT) with exponential forgetting (EF) has been proposed for time-varying systems to improve the data dependence of FRIT, which is a direct data-driven tuning method.
However, the EF-based method is not a reliable controller because it can cause significant degradation of the control performance and instability unless the persistent excitation (PE) condition is satisfied.
To solve this problem, we propose a new A-FRIT method based on directional forgetting (DF) and exponential resetting that can forget old data without instability regardless of the PE condition.
To confirm the effectiveness of the proposed method, we applied it to artificial muscle control with strong asymmetric hysteresis characteristics and evaluated its robust performance against load changes during the experiment.
The experimental results show that the proposed method based on DF achieves high control performance and is robust against changes in the characteristics and/or target trajectory.
The proposed method is also practical because it does not require system identification, model structure, or prior experimentation.
\end{abstract}

\section{INTRODUCTION}
In recent years, data-driven control has been actively researched because it does not explicitly use mathematical models, thus reducing typical time–consuming routines for model derivation, identification, and evaluation \cite{DDC1,DDC2}.
Thus, it is very effective for systems that require complex mathematical models and has been used in many applications.
For example, many methods have been proposed, such as virtual reference feedback tuning (VRFT) \cite{VRFT} and fictitious reference iterative tuning (FRIT) \cite{FRIT}, which are known as direct data-driven tuning methods, model-free adaptive control \cite{MFAC}, which are based on dynamic linearization techniques, reinforcement learning \cite{RL}, and neural network-based control \cite{NN}, which are classified as machine learning, and data-driven model predictive control based on Willems's fundamental lemma \cite{DDMPC}.

Among these, we focused on the direct data-driven controller tuning method, which is easy to apply because of its simple structure and few assumptions for practical applications.
FRIT and VRFT introduce the concept of virtual signals, allowing the controller parameters to be tuned using only single input-output (I/O) data from preliminary experiments.
The controller can be selected as PID, I-PD, or another relatively arbitrary controller.
Moreover, unlike VRFT, FRIT does not require a pre-filter to guarantee properness; thus, it is more practical method.

However, these methods are strongly dependent on prior experimental data, and assume a time-invariant system for plant systems.
Therefore, control performance is no longer guaranteed when the system characteristics change.
In addition, achieving the desired control performance is difficult when the target trajectory for the pre-experimental data differs from that of the target trajectory for control.
To solve this problem, an adaptive FRIT (A-FRIT) that tunes the controller online based on the recursive least squares (RLS) method has been proposed, and its effectiveness has been verified \cite{A-FRIT1,A-FRIT2}.
However, the A-FRIT algorithm is not sufficient to significantly improve the control performance compared to the normal FRIT, which is an offline optimization.
This is because the evaluation function for the RLS algorithm uses all I/O data from the past to the present without weighting for estimation.
To address this problem, it is common to introduce a forgetting factor in the field of identification and estimation.
Although the most common method of exponential forgetting (EF) was applied in \cite{A-FRIT2}, it does not provide sufficient forgetting because of the forgetting factor design, which is very close to $1$.
This is because lower forgetting factors for efficient forgetting may easily degrade or destabilize adaptive systems.
Such destabilization depends on the value of the regressor vector, or more specifically, on the persistent excitation (PE) condition.

In this paper, we propose a new robust A-FRIT method that achieves high control performance and robustness against characteristic changes and noise by sufficient forgetting, regardless of the PE condition.
A-FRIT methods and data-driven controls for time-varying systems in the absence of PE have rarely been discussed so far.
In this study, we apply a forgetting factor method that guarantees the boundedness of the information matrix independent of the value of the regressor vector.
The proposed method allows designers to select the value of the forgetting factor arbitrarily without instability. Therefore, the proposed method can perform appropriate forgetting for various practical applications, and is expected to achieve high robustness and control performance.
To verify the effectiveness of the proposed method, we applied it to a tap-water-driven artificial muscle \cite{HAM1} with strong asymmetric hysteresis characteristics.
Furthermore, it is difficult to design a model-based control system for the muscle because the shape of the hysteresis characteristic of this muscle changes significantly even when the load changes \cite{HAM2}.
Therefore, we verified the robustness of the proposed method by changing the operating point and load of the artificial muscle during the experiment and/or by different target trajectories during parameter tuning and control.


\section{Overview of FRIT}
\subsection{FRIT}
FRIT can determine the optimal controller parameters by solving a nonlinear optimization problem based on single-closed-loop I/O data without a mathematical model.
Let $z$ be the time shift operator and $k$ be the time step.
The objective of FRIT is to minimize the following model reference evaluation function:
\begin{equation}\label{ma:eva_FRIT1}
J(\theta)=\left\|\frac{G(z)C(\theta,z)}{1+G(z)C(\theta,z)}r(k)-G_m(z)r(k)\right\|^2_2,
\end{equation}
where $\|\cdot\|_2$ denotes the Euclidean norm, $G(z)$ denotes the plant system, $G_m(z)$ denotes the reference model appropriately designed by the designer, $C(\theta,z)\triangleq \beta^T(z)\theta$ denotes a controller with controller structure $\beta(z)\in\mathbb{R}^n$ and controller parameters $\theta\in\mathbb{R}^n$, and $r(k)\in\mathbb{R}$ denotes the target trajectory.
The controller parameters were determined numerically such that the closed-loop system matched the reference model.
However, because $G(z)$ in (\ref{ma:eva_FRIT1}) is assumed to be unknown, it cannot be solved.
Therefore, by using prior experimental data, FRIT rewrites the evaluation function as follows:
\begin{equation}\label{ma:eva_FRIT2}
J_{\mathrm{FRIT}}(\theta)=\sum_{k=1}^{N}\left[y_0(k)-G_m(z)\tilde{r}(\theta,k)\right]^2,
\end{equation}
where $y_0(k)\in\mathbb{R}$ represents the output data based on prior single closed-loop experimental data and $\tilde{r}\in\mathbb{R}$ denotes a fictitious reference signal, as follows:
\begin{equation}\label{ma:r_tilde}
\tilde{r}(\theta,k)=C^{-1}(z,\theta)u_0(k)+y_0(k)
\end{equation}
with the input data based on a prior single closed-loop experiment $u_0(k)\in\mathbb{R}$ and inverse controller $C^{-1}(z,\theta)$.
(\ref{ma:eva_FRIT2}) has no unknown values. Therefore, the controller parameters that minimize (\ref{ma:eva_FRIT2}) can be determined without using a mathematical model.

\begin{rem}
Note that the control performance using the obtained controller parameters is strongly dependent on the prior experimental data and initial controller parameters.
Therefore, it is important to introduce online tuning to compensate for changes in the optimal controller parameters due to changes in the characteristics or target trajectory.
\end{rem}

\subsection{A-FRIT}
Because FRIT has a nonlinear evaluation function, it requires a nonlinear optimization problem to be solved.
To update the controller parameters recursively, an evaluation function was constructed with a convex function concerning $\theta$ based on the results of Wakasa $et\ al.$ \cite{A-FRIT1}.
Consider an ideal situation in which the evaluation function is zero; that is, assume that there exists a $\theta$ such that the following relationship holds:
\begin{equation}\label{ma:ass}
y_0(k)-G_m(z)\tilde{r}(\theta,k)\equiv0.
\end{equation}
From (\ref{ma:r_tilde}) and (\ref{ma:ass}), we have
\begin{equation}\label{ma:AFRIT_temp}
C(\theta,z)y_0(k) = G_m(z)u_0(k)+C(\theta,z)G_m(z)y_0(k).
\end{equation}
The assumption that the evaluation function is equivalent to zero is not realistic. Thus, (\ref{ma:AFRIT_temp}) is redefined as the following auxiliary error:
\begin{equation}\label{ma:auxiliary error}
\hat{e}(k)\triangleq C(z)\{1-G_m(z)\}y_0(k)-G_m(z)u_0(k).
\end{equation}
The evaluation function is then rewritten as follows:
\begin{equation}\label{ma:eva_A-FRIT}
\hat{J}_{\mathrm{FRIT}}(\theta)=\sum_{k=1}^{N}\hat{e}^2(k)\triangleq\sum_{k=1}^{N}\left[\phi^T(k)\theta-d(k)\right]^2,
\end{equation}
where $\phi(k)\triangleq \beta(z)\{1-G_m(z)\}y_0(k)\in\mathbb{R}^n$ and $d(k)\triangleq G_m(z)u_0(k)\in\mathbb{R}$.
Because (\ref{ma:eva_A-FRIT}) is a convex function, it can be efficiently solved as a least-squares problem.
The parameter update law for minimizing the evaluation function above can be expressed as follows:
\begin{equation}\label{ma:RLS}
\left\{
\begin{aligned}
P(k) &= \frac{1}{\mu}\left[P(k-1)-\frac{P(k-1) \phi(k)\phi^T(k)P(k-1)}{\mu + \phi^T(k)P(k-1)\phi(k)}\right]\\
\hat{\theta}(k) &= \hat{\theta}(k-1) + P(k)\phi(k)\left[\phi^T(k)\hat{\theta}(k)-d(k)\right]
\end{aligned}
\right.
\end{equation}
where $P(k)\in\mathbb{R}^{n\times n}$ denotes the covariance matrix; $\phi(k)\in\mathbb{R}^{n}$ denotes the regressor vector; $\mu\in(0,1]$ denotes the forgetting factor; and $\hat{\theta}\in\mathbb{R}^n$ denotes the estimated controller parameters.


In this study, for simplicity, controller $C(\theta,z)$ is assumed to be a PID controller; that is, $\theta = [K_P,\ K_I,\ K_D]^T,\ \beta(z)=[1,\ {T_s}/{(1-z^{-1})},\ {(1-z^{-1})}/{T_s}]^T$
where $T_s$ denotes sampling time.
Using the controller parameters estimated by the updated law (\ref{ma:RLS}), the control input can be expressed as follows:
\begin{equation}
u(k)=C(\hat{\theta},z)e(k)=\hat{\theta}^T(k)\beta(z)e(k)
\end{equation}
where $e(k)\triangleq r(k)-y(k)\in\mathbb{R}$ denotes the tracking error with the reference trajectory $r(k)\in\mathbb{R}$. 

\section{FF-based A-FRIT}
\subsection{Forgetting factor}
It is well known that the introduction of forgetting factors improves robustness for time-varying systems \cite{FF1}.
Several types of forgetting factor algorithms have been proposed, primarily focusing on the boundedness of the information matrix and stability analysis \cite{FF2}.
In particular, we examine three typical forgetting factor algorithms: EF, directional forgetting (DF) \cite{DF}, and exponential resetting (ER) \cite{ER1,ER2}, which is an extension of EF, and verify their effectiveness by combining them with the A-FRIT algorithm.
We consider the information matrix, which is defined as the inverse of the covariance matrix.

\subsubsection{Exponential forgetting}
EF is the most common forgetting method for uniformly forgetting past data.
The updated law of the information matrix for EF can be expressed as follows:
\begin{equation}
R(k) = \mu R(k-1) + \phi(k)\phi^T(k),
\end{equation}
where $R(k)\in\mathbb{R}^{n\times n}$ is the information matrix.
If the PE condition is satisfied, then the existence of positive constant upper and lower bounds for this information matrix is guaranteed, i.e., $^\exists\alpha,\beta>0\ \mathrm{s.t.}\ \alpha I\leq R(k)\leq\beta I,\ ^\forall k\geq 0$.
Otherwise, EF cannot guarantee the existence of a lower positive constant value.
Hence, the covariance matrix $P(k)$ acts as an update gain and implies that the inverse of the information matrix may diverge, because the positive definiteness of the information matrix can no longer be guaranteed.
This phenomenon is known as an estimator windup \cite{DF}.
Even if the inverse of the covariance matrix does not diverge, it is not a robust estimation because it can be strongly affected by noise.

\subsubsection{Directional forgetting}
To avoid estimator windup, DF algorithms were proposed \cite{DF}.
Several types of DF algorithms have been proposed, including those proposed by Liyu and Schwartz \cite{DF}.
The update law of the information matrix for DF can be decomposed as follows:
\begin{equation}
R(k) = R_1(k-1) + \mu R_2(k-1) + \phi(k)\phi^T(k),
\end{equation}
where the positive semidefinite matrix $R_1(k-1)\in\mathbb{R}^{n\times n}\ \mathrm{s.t.}\ R_1(k-1)\phi(k)=0,\ \phi(k)\neq 0$ and its rank is $n-1$, and positive semidefinite matrix $R_2(k-1)\in\mathbb{R}^{n\times n}\ \mathrm{s.t.}\ R_2(k-1)\phi(k)=R(k-1)\phi(k)$.
Thus, by orthogonally decomposing the information matrix $R(k-1)$ into $R_1(k-1)$ and $R_2(k-1)$, the DF algorithm forgets only $R_2(k-1)$, which has rank $1$.
The forgotten values of the information matrix $R_2(k-1)$ are complemented by $\phi(k)\phi^T(k)$, and the information matrix $R(k)$ always maintains positive definiteness regardless of the PE condition.

\begin{rem}
The combination of decomposing $R_1(k-1)$ and $R_2(k-1)$ is generally not unique.
However, it has been shown that $R_2(k-1)$ can be uniquely obtained provided that its rank is $1$ using lemma 1 in \cite{DF}.
\end{rem}

\subsubsection{Exponential resetting}
The ER algorithm was proposed by \cite{ER1} and generalized by \cite{ER2} to improve the performance of EF.
The updated law of the information matrix for ER can be expressed
as follows
\begin{equation}
R(k) = \mu R(k-1) + (1-\mu)R_{\infty} + \phi(k)\phi^T(k)
\end{equation}
where $R_{\infty}\in\mathbb{R}^{n\times n}\ \mathrm{s.t.}\ \lim_{k\to\infty}R(k)=R_{\infty}$ denotes an ideal information matrix under no PE condition.
Note that $R_{\infty}$ satisfies positive definiteness and is selected such that $R(0)\geq R_{\infty}$ with the initial information matrix $R(0)$.
The ER algorithm is a recently developed method that maintains parameter convergence while suppressing the estimated windup when the PE condition is not satisfied.
If the PE condition is satisfied, fast parameter convergence can be expected, as in EF.

\subsection{Proposed methods}
Two A-FRIT algorithms based on DF and ER are presented.
This is the first time that the proposed method has been combined with A-FRIT.

\subsubsection{DF-based A-FRIT algorithm}
The A-FRIT algorithm based on DF can be expressed as follows:



\begin{equation}
\left\{
\begin{aligned}
\hat{\theta}(k)&=\hat{\theta}(k-1)+P(k)\phi(k)[d(k)-\phi^T(k)\hat{\theta}(k-1)]\\
P(k)&=\bar{P}(k-1)-\frac{\bar{P}(k-1)\phi(k)\phi^T(k)\bar{P}(k-1)}{\phi^T(k)\bar{P}(k-1)\phi(k)}\\
\bar{P}(k-1)&=\left\{
\begin{aligned}
&P(k-1)+\frac{1-\mu}{\mu}\frac{\phi(k)\phi^T(k)}{\phi^T(k)R(k-1)\phi(k)},\\  &\qquad\qquad\qquad\qquad\qquad\qquad\quad\|\phi(k)\|>\varepsilon\\
&P(k-1),\quad \|\phi(k)\|\leq\varepsilon
\end{aligned}
\right.\\
R(k)&=[I-M(k)]R(k-1)+\phi(k)\phi^T(k)\\
M(k)&=\left\{
\begin{aligned}
&(1-\mu)\frac{R(k-1)\phi(k)\phi^T(k)}{\phi^T(k)R(k-1)\phi(k)},\quad\ \|\phi(k)\|>\varepsilon\\
&0,\quad \|\phi(k)\|\leq\varepsilon
\end{aligned}
\right.
\end{aligned}
\right.
\end{equation}
where scalar $\varepsilon >0$ denotes the design parameter, which is determined by the noise signal-to-noise ratio, and works as a deadzone for updating.

\subsubsection{ER-based A-FRIT algorithm}
The A-FRIT algorithm based on ER can be expressed as follows:

\begin{equation}
\left\{
\begin{aligned}
\hat{\theta}(k)&=\hat{\theta}(k-1)+P(k)\phi(k)[d(k)-\phi^T(k)\hat{\theta}(k-1)]\\
R(k)&= \mu R(k-1)+ (1-\mu) R_{\infty}+\phi(k)\phi^T(k)\\
P(k)&= R^{-1}(k)
\end{aligned}
\right.
\end{equation}

\begin{figure}[h]
\centering
\includegraphics[width=8.5cm]{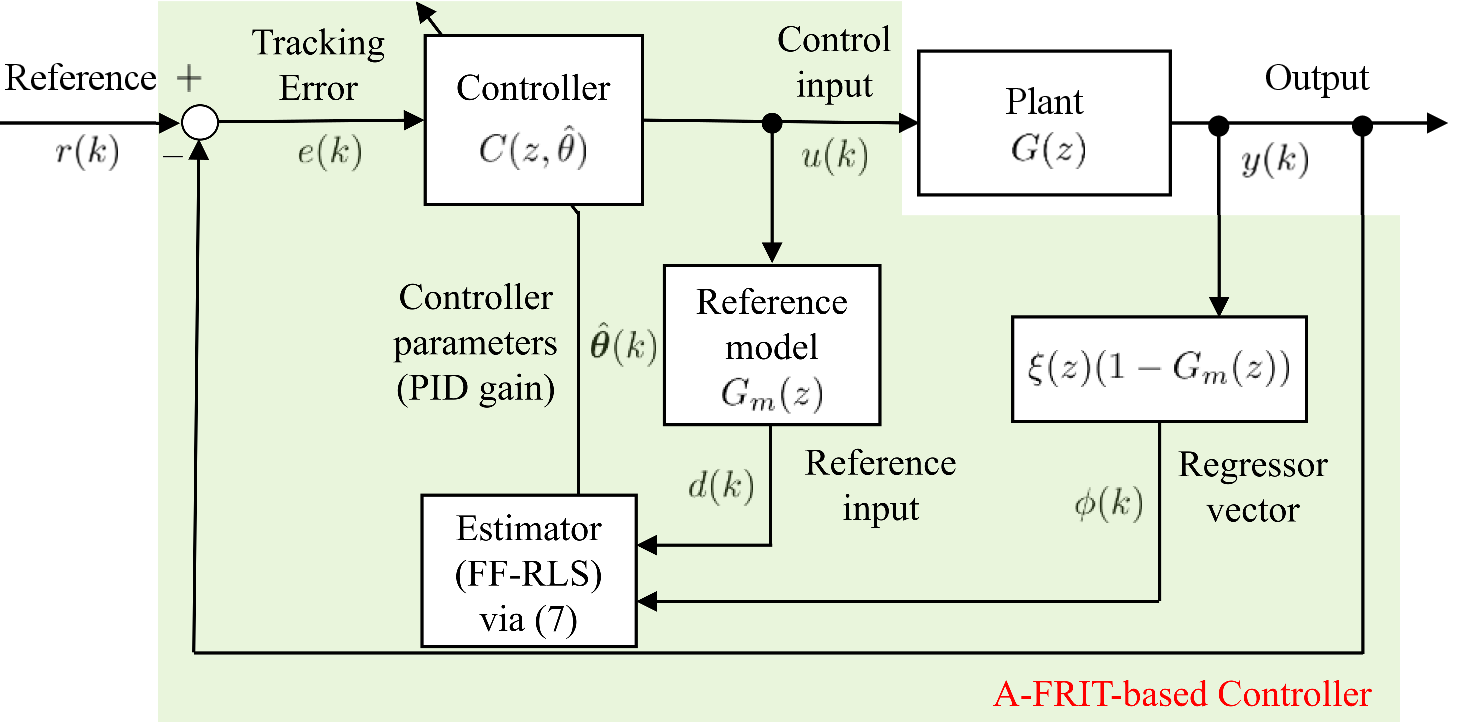}
  \caption{Block diagram of the adaptive FRIT algorithm}
  \label{fig:A-FRIT}
\end{figure}

Figure \ref{fig:A-FRIT} presents a block diagram of the proposed method.
As described in (\ref{ma:eva_A-FRIT}) in Section $\mathrm{I}\hspace{-1.2pt}\mathrm{I}$, the reference input and regressor vectors were calculated from the I/O data, and the above estimation algorithm was applied.
The proposed methods are expected to select a sufficiently small value for the forgetting factor, and thus are expected to be effective data-driven control methods for time-varying systems.

\section{Experimental results and discussion}
To verify the effectiveness of the proposed methods, they were applied to an artificial muscle control system and evaluated based on the following three aspects.
1) Several methods were compared to evaluate the best method for a staircase target trajectory by changing the driving point.
Moreover, the effect of the initial PID gain is discussed.
2) For DF-based A-FRIT, the best values were evaluated experimentally by changing the forgetting factor value.
3) The robustness of the proposed methods was evaluated when the load changed during the experiment, thus validating the effectiveness of the time-varying systems.

\begin{figure}[h]
\begin{minipage}[b]{0.49\linewidth}
\centering
\includegraphics[width=4.5cm]{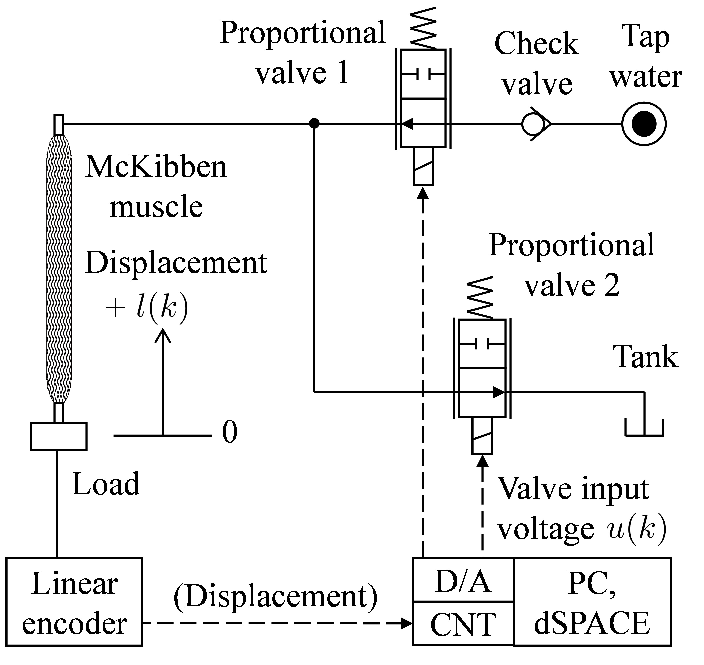}
  \subcaption{Experimental circuit}
  \end{minipage}
 \begin{minipage}[b]{0.49\linewidth}
 \centering
 \includegraphics[width=3cm]{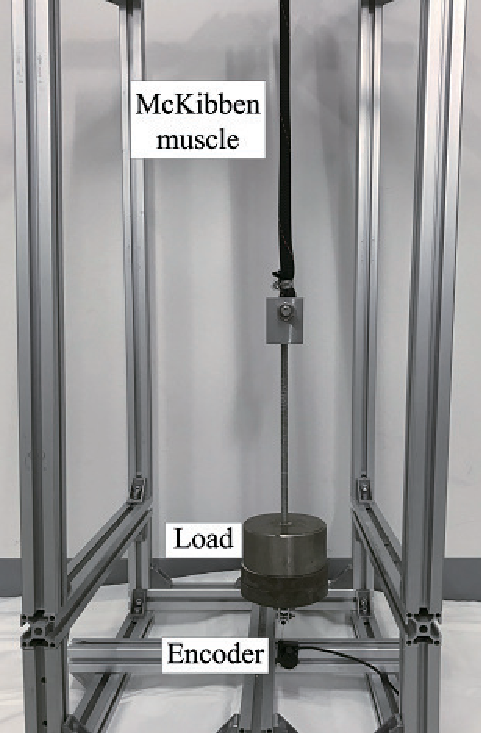}
 \subcaption{Experimental equipment}
 \end{minipage}
 \caption{Experimental setup}
  \label{fig:setup}
\end{figure}

Figure \ref{fig:setup} illustrates the experimental circuit and equipment.
In this study, we used a tap-water-driven handmade artificial muscle with a natural length of $360$ mm \cite{HAM1,HAM2}.

\subsubsection{Comparison of FF-based A-FRIT algorithm}
We compared several methods including FRIT, A-FRIT, EF-based A-FRIT, DF-based A-FRIT, and ER-based A-FRIT.
The design parameters in this experiment were set to the initial covariance matrix $P(0)=100I_3$, initial information matrix $R(0)=0.01I_3$, ideal information matrix $R_{\infty}=0.01I_3$, threshold value $\varepsilon=10^{-3}$, and forgetting factor $\mu=0.99$.
The reference model $G_m(z)$ was set to $0.0095/(z-0.99)$ as a discrete-time transfer function with a time constant of 1 s, gain of 1, and sampling time of 0.01 s.
Three initial PID gains were compared: with and without offline tuning, and the result of tuning for different target trajectories, which is frequency 0.3 Hz, amplitude 30 m, and offset 20 mm.
The initial PID gain for tuning was set to $\hat{\theta}(0)=[0.1,\ 0.1,\ 0.01]^T$, and the PID gain for same target trajectory and the PID gain for different target trajectory were optimized offline using FRIT as $\hat{\theta}(0)=[0.107,\ 0.1515,\ 0.0115]^T$, $\hat{\theta}(0)=[0.0828,\ 0.2737,\ 0.0539]^T$, respectively.
Note that FRIT is performed using a fixed gain with an initial PID gain.

\begin{figure}[h]
\centering
\includegraphics[width=8.5cm]{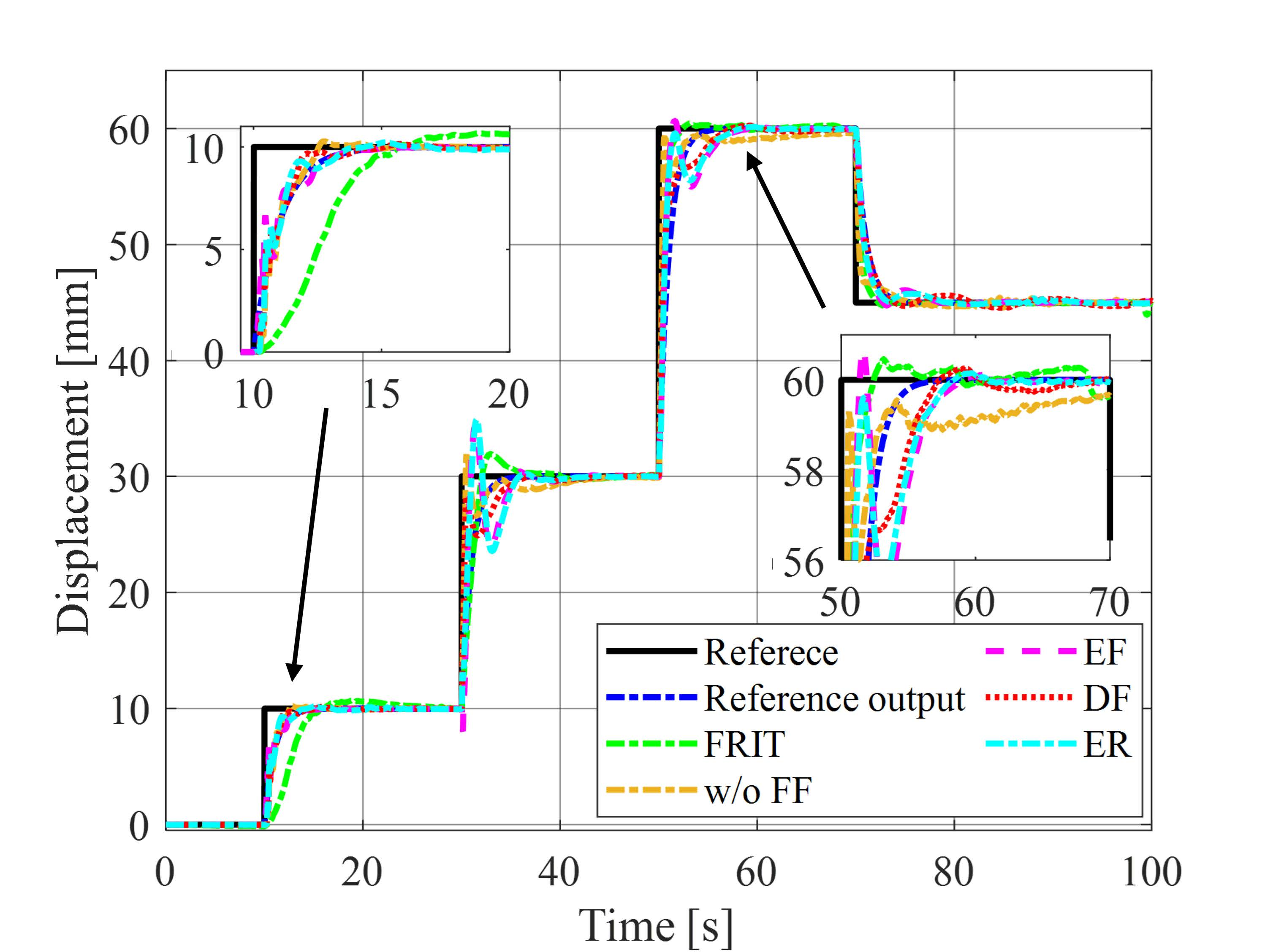}
  \caption{Comparative control performance of FF-based A-FRIT with forgetting factor $\mu=0.99$ for staircase square reference}
  \label{fig:comp_FF}
\end{figure}

\begin{figure}[h]
\centering
\includegraphics[width=8.5cm]{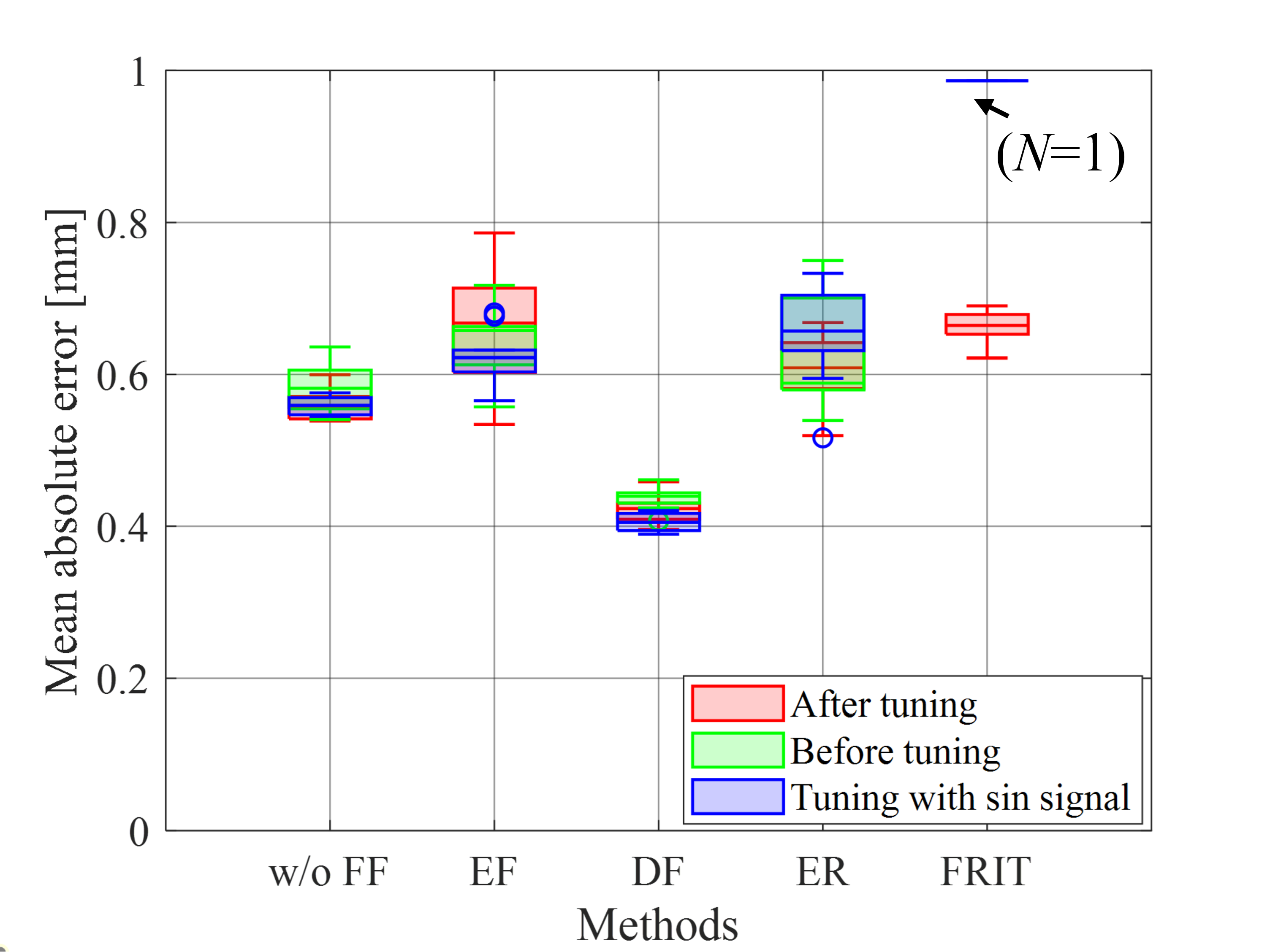}
  \caption{Evaluation results using box plots based on the MAE of the tracking error between the reference output and displacement before and after controller parameter tuning (number of each trial experiment: 10 times)}
  \label{fig:box_comp_FF}
\end{figure}

Figure \ref{fig:comp_FF} compares the control performances of several methods.
The FRIT algorithm provides PID gains that minimize the evaluation function (\ref{ma:eva_FRIT2}) for all intervals, which degrades the control performance during the transient performance.
The A-FRIT algorithm updates the PID gains recursively, resulting in high transient performance.
Moreover, focusing on approximately $50$–$70$ s, the method without FF did not track the target trajectory.
This is because they continued to have all past data, indicating the need to introduce these forgetting factors.
In contrast, FF-based A-FRIT algorithms, such as EF, DF, and ER, achieved better control performance in the same intervals. 
However, the EF and ER-based algorithms became highly oscillatory at approximately $30$ or $60$ s, owing to disturbances or noise.
In contrast, the DF-based algorithm exhibited superior robustness.
These results are influenced by the behavior of the covariance matrix, which is discussed in 3) below.

Next, Fig. \ref{fig:box_comp_FF} shows the box plots of the evaluation results based on the mean absolute error (MAE) of the tracking error between the reference output and displacement.
In particular, the EF- and ER-based algorithms exhibited a large variation in control performance owing to their wide interquartile ranges.
However, the proposed DF-based algorithm has a very narrow range and achieves the highest control performance among all methods.
In addition, the FRIT algorithm deteriorates significantly when the PID gain is tuned using different from the target trajectory for control.
Note that the number of samples in this experiment is only one because the artificial muscle becomes so oscillatory that it is virtually uncontrollable.
On the other hand, the A-FRIT algorithms were independent of the initial PID gain, and, in particular, the DF-based A-FRIT achieved high robustness.
Also, this implies that the FF-based A-FRIT does not require prior experimental data; thus, we can reduce time–consuming routines for controller parameter tuning. 

\subsubsection{Effect of forgetting factor value for DF-based A-FRIT}
For the DF-based algorithm, we experimentally confirmed the effect of the forgetting factor on the control performance.
In this study, five forgetting factor values were applied: $0.99$, $0.90$, $0.85$, $0.80$, and $0.75$.
The other experimental conditions were the same as those described in 1) above.

\begin{figure}[t]
\centering
\includegraphics[width=8.5cm]{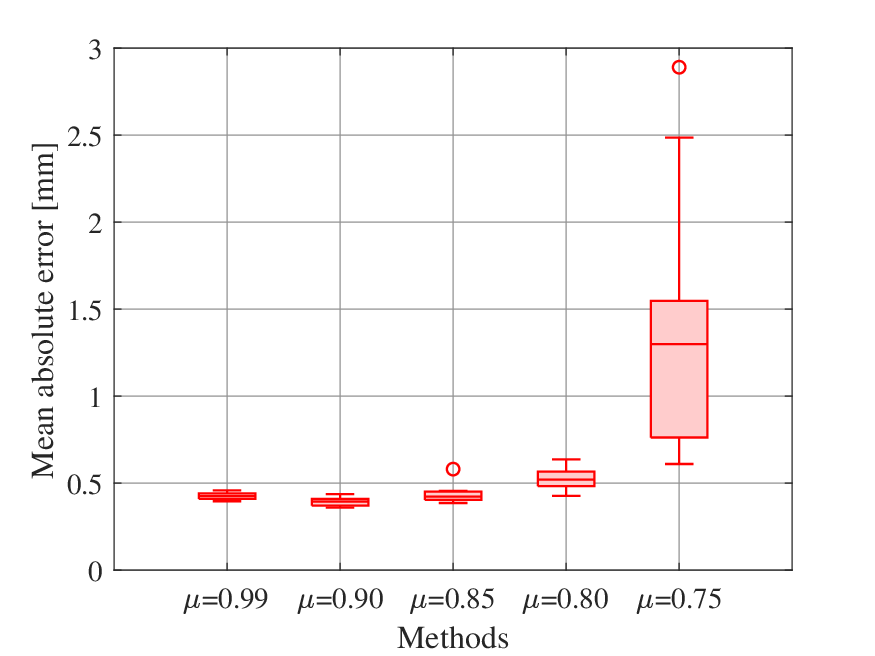}
  \caption{Evaluation results for DF-based A-FRIT using box plots based on the MAE of the tracking error between the reference output and displacement by several forgetting factor values (number of each trial experiment: 10 times)}
  \label{fig:box_DF}
\end{figure}

Figure \ref{fig:box_DF} shows box plots using the evaluation results based on the MAE of the DF-based A-FRIT algorithm. 
This figure shows that the value of $0.9$ was the best forgetting factor among $5$ values.
The closed-loop system did not become unstable even when the value of the forgetting factor was set to $0.75$.
In general, the EF algorithm cannot sufficiently reduce the value of the forgetting factor; thus, we must select a value as close to 1 as possible, which cannot be treated as effectively as a design parameter.
The control performances of EF- and ER-based algorithms become oscillatory and eventually unstable when the value is less than 0.99.
Therefore, the forgetting factor of DF-based A-FRIT can be adjusted as a design parameter because it can be selected arbitrarily for efficient forgetting.

\subsubsection{Robustness to characteristic changes}
To evaluate the robustness to changes in the characteristics, the load on the artificial muscle was manually changed from $44$ N to $68$ N at approximately 50 s during the experiments.
The forgetting factors for each method were set to $0.99$ for EF, $0.9$ for DF, and $0.99$ for ER to obtain the best control performance.
For simplicity, the target trajectory was set to a constant value of $50$ mm.
The initial value of the PID gain $\theta(0)=[0.162,\ 0.129,\ 0.061]^T$ was optimized by FRIT from the prior I/O data of the step signal.
The other conditions were the same as those described in 1).
In the experiments, we focused on the transient response after the load change.

\begin{figure}[!t]
\centering
\includegraphics[width=8.5cm]{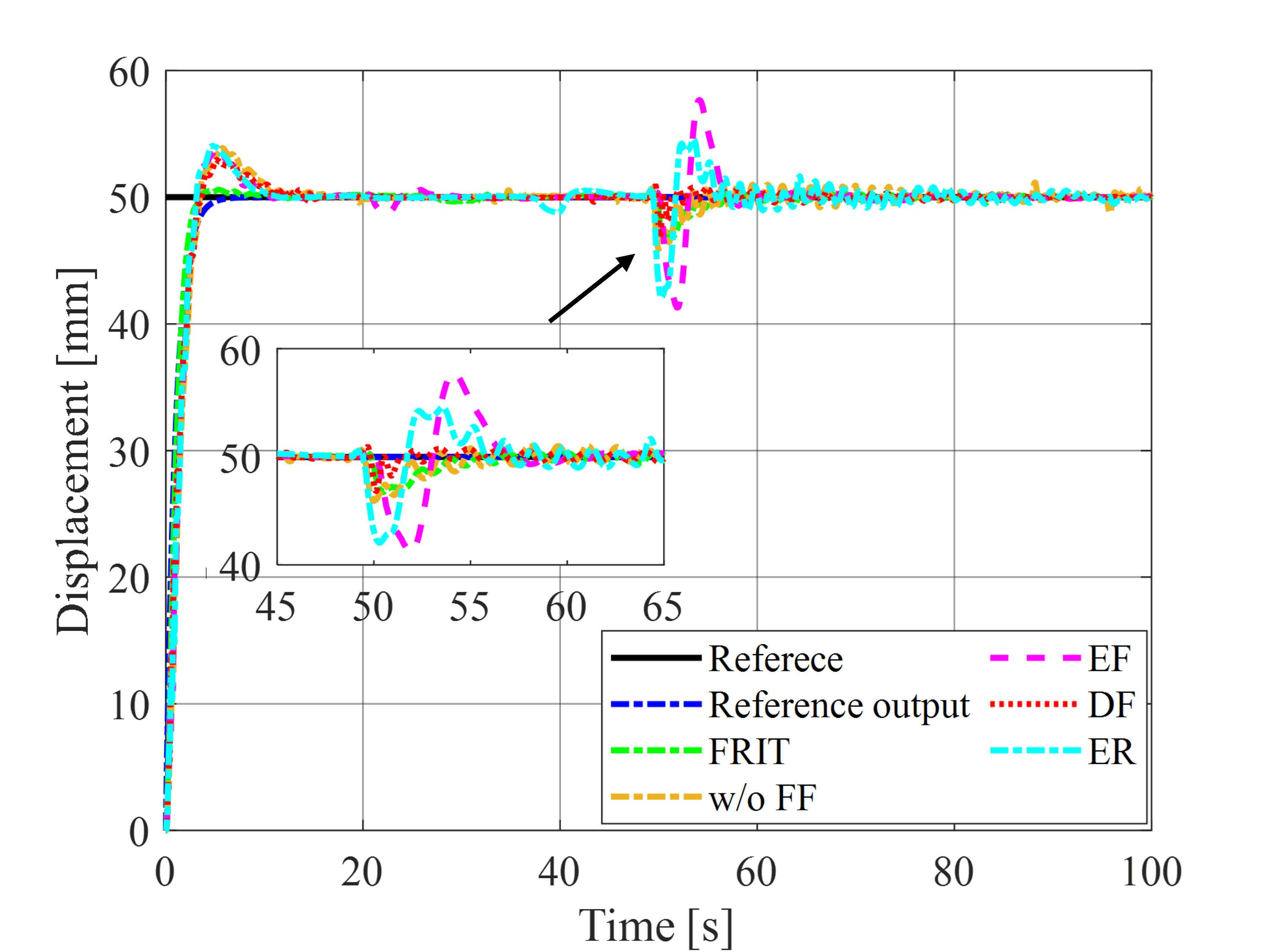}
  \caption{Comparison of control performance for FF-based A-FRIT with forgetting factor value (EF, ER: $\mu=0.99$, DF $\mu=0.9$) under load change at around 50 s}
  \label{fig:control performance_LC}
\end{figure}

\begin{figure}[!t]
\centering
\begin{minipage}[b]{0.48\linewidth}
    \centering
    \includegraphics[keepaspectratio, scale=0.4]{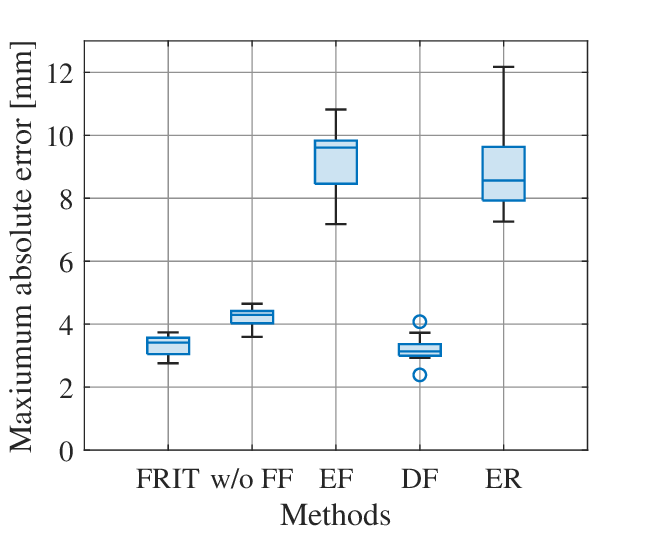} 
    \subcaption{Maximum absolute error}
  \end{minipage}
\begin{minipage}[b]{0.48\linewidth}
    \centering
    \includegraphics[keepaspectratio, scale=0.4]{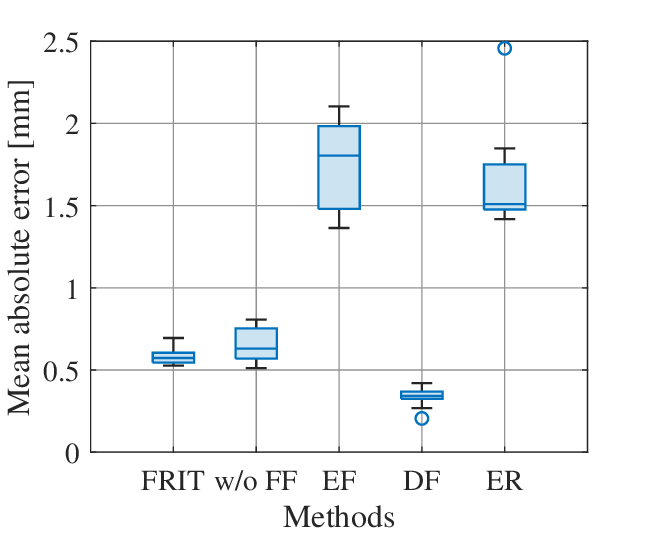}
    \subcaption{Mean absolute error}
  \end{minipage}
  \caption{Evaluation results using box plots around load change (evaluation interval is 45-65 s, number of each trial experiment: 10 times)}
  \label{fig:boxplot_3}
\end{figure}

\begin{figure}[!t]
\centering
\includegraphics[width=8.5cm]{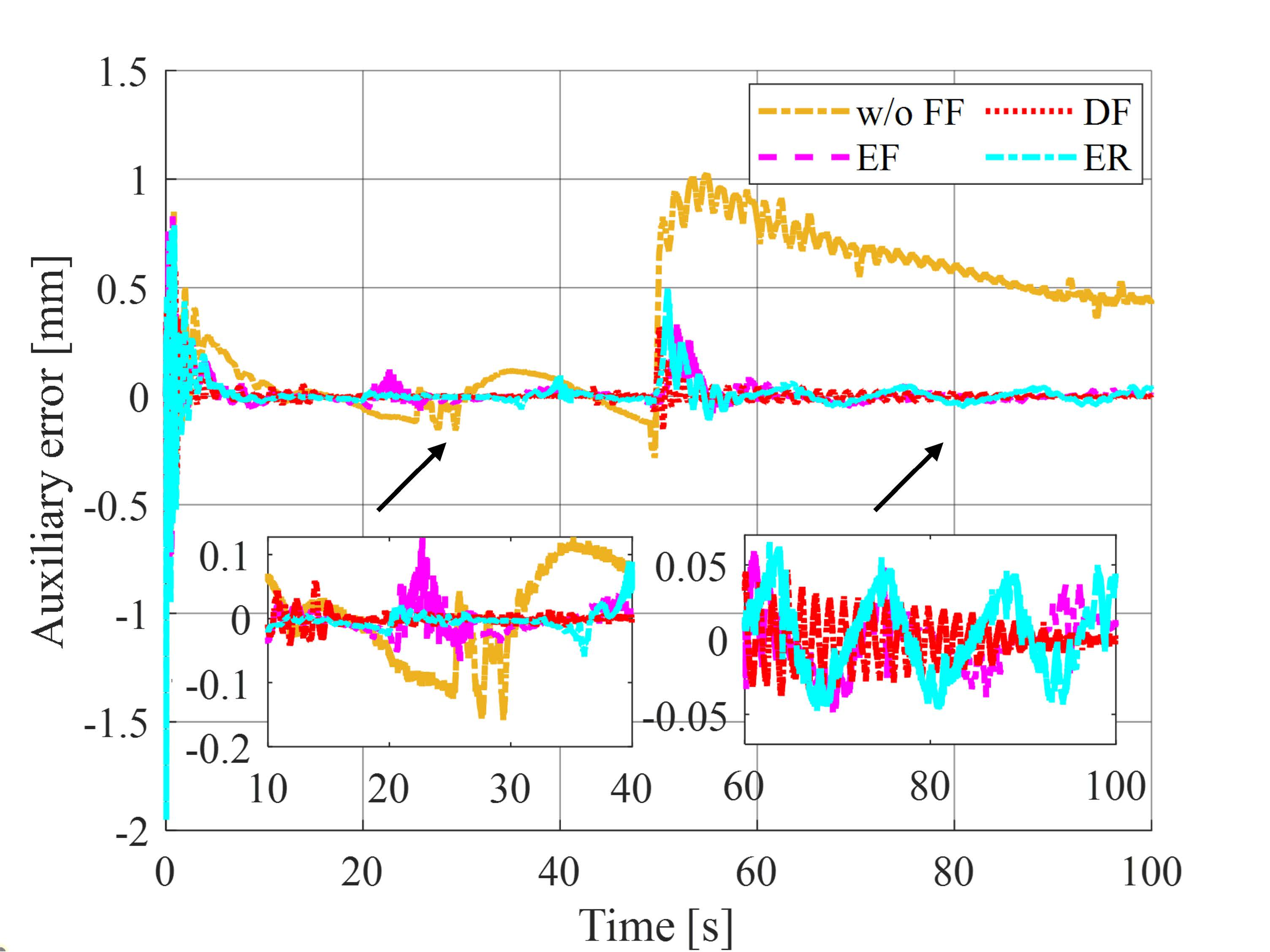}
  \caption{Comparison of auxiliary error $\hat{e}(k)$ (\ref{ma:auxiliary error}) under load change}
  \label{fig:auxiliary error}
\end{figure}

\begin{figure}[!t]
\centering
\begin{minipage}[b]{0.48\linewidth}
    \centering
    \includegraphics[keepaspectratio, scale=0.09]{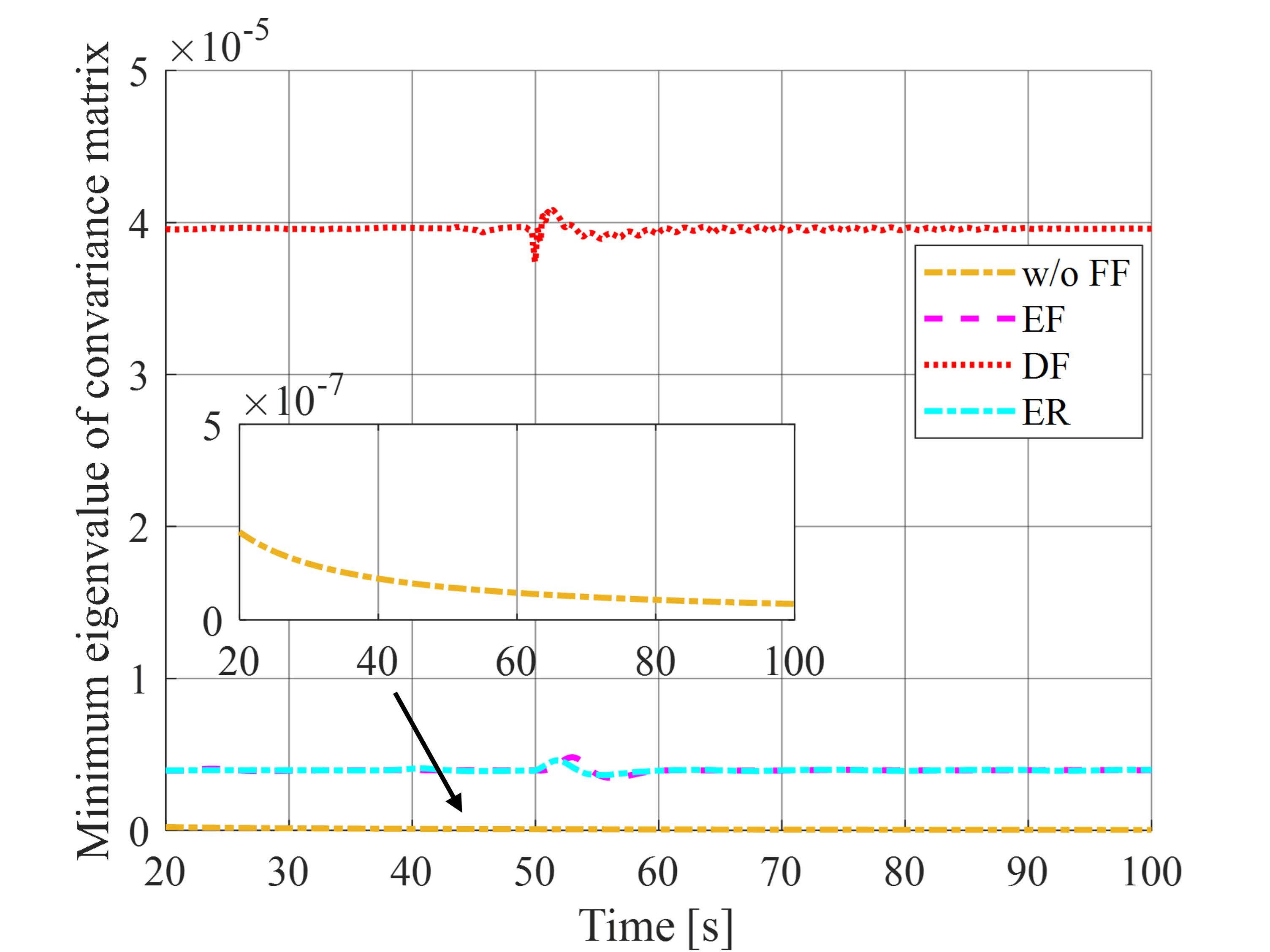} 
    \subcaption{Minimum eigenvalue}
  \end{minipage}
\begin{minipage}[b]{0.48\linewidth}
    \centering
    \includegraphics[keepaspectratio, scale=0.09]{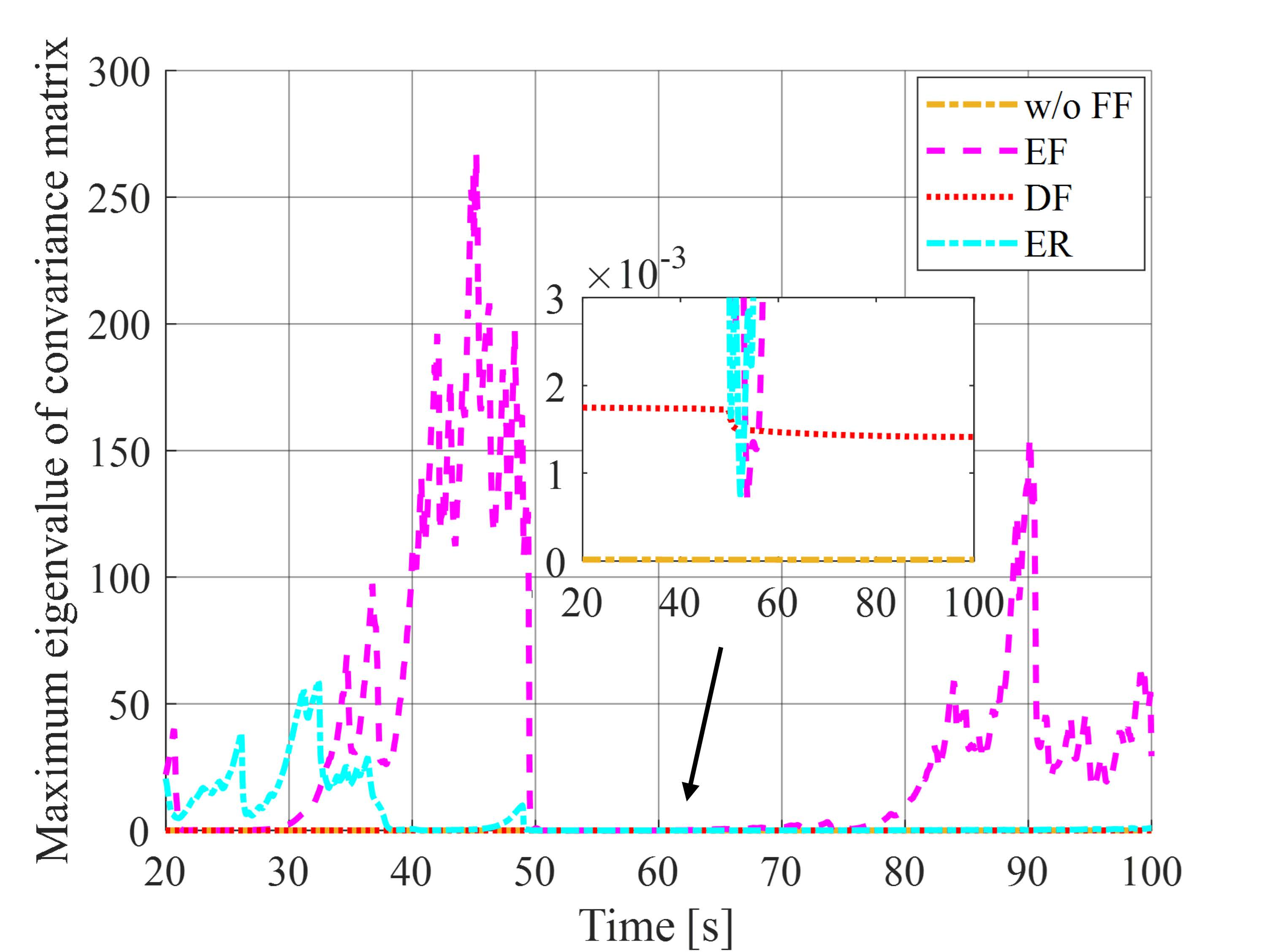}
    \subcaption{Maxiumum eigenvalue}
  \end{minipage}
  \caption{Comparison of the eigenvalue of the covariance matrix $P(k)$ for the FF-based A-FRIT algorithm under load change in the steady-state response}
  \label{fig:covariance}
\end{figure}

Figures \ref{fig:control performance_LC} and \ref{fig:boxplot_3} compare the control performance, including changing the load, and its evaluation using the maximum absolute error (maxAE) and MAE during the load change interval, respectively.
The maxAE results indicate that the DF-based method has the same level of degradation as FRIT, which is controlled by a fixed gain.
In general, introducing adaptive systems often causes significant transient performance degradation owing to characteristic changes or noise; however, this is not the case with the proposed DF-based method.
From the MAE results, the DF-based method exhibited the best control performance, that is it could quickly return to the target trajectory.
This is because the DF-based method can reduce the auxiliary error (\ref{ma:auxiliary error}), as shown in Fig. \ref{fig:auxiliary error}, thus, the accuracy of the controller parameter's estimation for the DF-based method is the highest.
Furthermore, Fig. \ref{fig:covariance} shows the eigenvalues of the covariance matrix in steady-state response.
The estimation accuracy of the w/o FF-based method did not improve after the load change because the minimum eigenvalue of the covariance matrix for the w/o FF-based method became extremely small.
However, the DF-based method has the largest minimum eigenvalue in comparison because we can set an effective forgetting factor value closer to $0$ without destabilizing the system; that is, it is more robust to time-varying systems.
Moreover, the maximum eigenvalues of EF- and ER-based methods are very large.
For the A-FRIT algorithm, it is difficult to satisfy the PE condition for the regressor vector owing to the adaptive control system.
The EF-based method does not guarantee positive definiteness of the information matrix when the PE condition is not satisfied, resulting in a larger maximum eigenvalue of the covariance matrix.
By contrast, the ER-based method can guarantee the positive definiteness of the information matrix regardless of the PE condition, but the resulting minimum eigenvalue of the information matrix is smaller.
This is because to guarantee the stability of the ER, the ideal information matrix $R_\infty$ must be chosen to satisfy $R(0)\geq R_\infty$
This implies a tradeoff between the adaptive capacities of the transient and steady-state responses.
Hence, it is difficult to apply an ER-based method to a practical adaptive control system that differs from system identification.
However, the DF-based method guarantees positive definiteness regardless of the PE condition, and both the small maximum and large minimum eigenvalues indicate that the condition number is very small.
Thus, the DF method achieves high robustness against characteristic changes.

\section{CONCLUSIONS}
A novel robust data-driven recursive controller design method is proposed that combines an adaptive FRIT algorithm with a forgetting factor, which guarantees positive definiteness under non-PE conditions based on DF and ER.
To verify the effectiveness of the proposed method, it was applied to a tap water-driven artificial muscle with strong asymmetric hysteresis characteristics.
The experimental results indicated that the DF-based method achieved higher robustness than the conventional method for all target trajectories, initial controller parameters, and load changes.
In addition, compared with the conventional method, the DF method did not cause instability even when the value of the forgetting factor was reduced to some extent, allowing the designer to make adjustments with confidence and ease.





\end{document}